\documentclass[conference]{IEEEtran}
\usepackage{amsmath,amssymb}
\usepackage{latexsym}
\usepackage{CJK}
\usepackage{cite}   
\usepackage{algorithm}
\usepackage{algorithmic}
\usepackage{caption}
\usepackage{multirow}
\usepackage{amsthm}
\usepackage{amsmath,amssymb}
\usepackage{xcolor}

\usepackage{graphics}
\usepackage{graphicx}
\usepackage{epsfig}
\usepackage{graphicx}
\usepackage{epstopdf}
\usepackage{verbatim}
\usepackage{rotating}
\usepackage{booktabs}
\usepackage{setspace}
\usepackage{indentfirst}
\usepackage{lettrine}   
\usepackage[numbers]{natbib}
\usepackage{multicol}
\usepackage{cite}   

\usepackage{verbatim}  
\usepackage{graphicx}  
\usepackage{mathrsfs} 
\usepackage{algorithm} 
\usepackage{algorithmic} 
\usepackage{booktabs}
\usepackage{textcomp}  
\usepackage{multirow}  
\usepackage{lettrine}   
\usepackage{graphicx}  
\usepackage{bm}  
\usepackage{amsthm} 

\usepackage{subfigure} 

\newcommand{\myvec}[1]%
   {\stackrel{\raisebox{-2pt}[0pt][0pt]{\small$\rightharpoonup$}}{#1}}

\setcitestyle{open={},close={}}

\ifCLASSINFOpdf
\else
\fi

\begin{document}
%
\title{Shannon Shakes Hands with Chernoff: Big Data Viewpoint On Channel Information Measures}
%
%
%

\author{Shanyun Liu, ~Rui~She, Jiaxun Lu, ~Pingyi~Fan\\

\small
Tsinghua National Laboratory for Information Science and Technology(TNList),\\
Department of Electronic Engineering, Tsinghua University, Beijing, P.R. China\\
E-mail: liushany16@mails.tsinghua.edu.cn, sher15@mails.tsinghua.edu.cn, lujx14@mails.tsinghua.edu.cn, ~fpy@tsinghua.edu.cn}

\maketitle

\begin{abstract}
Shannon entropy is the most crucial foundation of Information Theory, which has been proven to be effective in many fields such as communications. R{\'{e}}nyi entropy and Chernoff information are other two popular measures of information with wide applications. The mutual information is effective to measure the channel information for the fact that it reflects the relation between output variables and input variables. In this paper, we reexamine these channel information measures in big data viewpoint by means of ACE algorithm. The simulated results show us that decomposition results of Shannon and Chernoff mutual information with respect to channel parameters are almost the same. In this sense, Shannon shakes hands with Chernoff since they are different measures of the same information quantity. We also propose a conjecture that there is nature of channel information which is only decided by the channel parameters.

\end{abstract}


\IEEEpeerreviewmaketitle

\section{Introduction}
Shannon entropy is the most crucial foundation of Information Theory. In this traditional Information Theory and its applications, relative entropy (or Kullback-Leibler distance) is the basic information divergence. Shannon Information Theory is useful for us to develop communications, data science and other subjects about information [\cite{verdu1998fifty}]. Due to its success, there are a lot of literature which attempts to advance these concepts. Unfortunately, nearly none of them have been widely adopted except the the R{\'{e}}nyi entropy [\cite{rrnyi1961measures}][\cite{van2014renyi}]. The R{\'{e}}nyi entropy has a wide range of applications. For example, according to [\cite{principe2000information}], it is useful to adopt R{\'{e}}nyi entropy in the traditional signal processing. Besides that, the R{\'{e}}nyi entropy was also used in independent component analysis (ICA) with the fact that the information measures based on R{\'{e}}nyi entropy could provide the distance measures among a cluster of probability densities [\cite{bao2003renyi}]. Moreover, some studies suggested that it could help us to do blind source separation (BSS) [\cite{hild2001blind}][\cite{hild2006analysis}]. The Chernoff information was developed to measure the information based on R{\'{e}}nyi entropy [\cite{van2014renyi}]. In fact, the Chernoff information is the exponent when minimizing the overall probability of error, so it is very useful in hypothesis testing [\cite{Elements}].\\ 

In the last 68 years of Information Theory development, a large amount of concepts have been provided to describe the the channel information. Among them, the most popular one is channel capacity. As a matter of fact, the channel capacity is defined as maximization mutual information. The mutual information is a measure of the amount of information and it is the reduction in uncertainty of one random variable when given the other variable [\cite{Elements}]. With respect to (w.r.t.) channel capacity, the mutual information is easier to use and expand, without loss of the ability to measure the channel "information". As a result, in order to be promoted to other non-Shannon entropy, the mutual information between channel input and output is adopted in this paper.\\ 

Unlike traditional research method of Information Theory, we attempt to revisit the traditional problem in big data viewpoint. Obviously, the era of big data is coming. Tha large amount of data can be easily gotten and the data processing is more and more important for us. Two data rules are believed. For one thing, the data does not lie. For other thing, we only need to guarantee that our conclusion is correct with very large probability rather than one in the era of big data [\cite{vapnik2006estimation}]. In this paper, we choose the alternating conditional expectation (ACE) algorithm as the tool to deal with data.\\

The ACE algorithm was proposed by Breiman and Friedman (1985) [\cite{breiman1985estimating}] for estimating the transformations of dependent variable and a set of independent variables in multiple regression that  estimate maximal correlation among the these variables [\cite{wang2004estimating}]. It can help us analyze multivariate function for it enabling multiple variable separation. Its effectiveness and correctness were provided in [\cite{breiman1985estimating}]. With the ACE algorithm, it is easy to separate the effect of different channel parameters so as to find the nature connection between channel information and the channel parameters. Furthermore, one can investigate the relation between Shannon and Chernoff as well.\\

The main contribution of this paper can be summarized as follows. We decompose the channel mutual information of Shannon and Chernoff w.r.t channel parameters by ACE algorithm. The simulated results show us that Shannon shakes hands with Chernoff in big data viewpoint. Based on these results, we put forward a conjecture that there is nature of channel information and no matter Shannon mutual information, Chernoff mutual information or other information measures, which are different measures of the same information quantity. This conclusion can help us to construct new information measures or  judge a new channel information measure is reasonable or not. \\

\subsection{Introduction of mutual Information}
\subsubsection{Shannon mutual Information}
The mutual information is a measure of the amount of information that one random variable contains about another variable. In fact, it describes the reduction in the uncertainty of one random variable if another one is known [\cite{Elements}].
According to [\cite{Elements}], the Shannon mutual information is defined as:
\begin{equation}\label{Shannon Information}
\begin{aligned}
  & I_S(X,Y) = D(p(x,y)||p(x)p(y))  \cr
  &  = \sum\limits_{x,y} {p(x,y)log{{p(x,y)} \over {p(x)p(y)}}}  \cr
 \end{aligned}
\end{equation}
where $X,Y$ are two random variables with marginal probability mass functions $p(x)$ and $p(y)$ and joint probability mass function $p(x,y)$. It is noted that $I(X;Y)$ is the relative entropy (or Kullback-leibler distance) between the joint distribution $p(x,y)$ and the product distribution $p(x)p(y)$. The relative entropy is given by [\cite{Elements}]:
\begin{equation}\label{KL distance}
D(p||q) = \sum\limits_{x \in \textit{X}} {p(x)log{{p(x)} \over {q(x)}}}
\end{equation}

\subsubsection{Chernoff mutual Information}
The Chernoff information is derived from the problem of classic hypothesis testing. Chernoff Information is the resulting error exponent when minimizing the overall probability of error [\cite{Elements}]. As defined in [\cite{Elements}], it is given by
\begin{equation}\label{Chernoff Information1}
\begin{aligned}
  I_C({{\rm{P}}_1},{P_2}) \buildrel \Delta \over =  - \mathop {\min }\limits_{0 \le \alpha  \le 1} \log \left( {\sum\limits_x {P_1^\alpha (x)P_2^{1 - \alpha }(x)} } \right)
 \end{aligned}
\end{equation}
where $\alpha$ is real parameter with $0\leq \alpha \leq 1$.

With the aid of Chernoff information to describe the relationship between the two variables, a new mutual information is defined as
\begin{equation}\label{Chernoff Information2}
\begin{aligned}
  I_C(X,Y) \buildrel \Delta \over =  - \mathop {\min }\limits_{0 \le \alpha  \le 1} \log \left( {\sum\limits_x {\sum\limits_y {{p^\alpha }(x,y){{(p(x)p(y))}^{1 - \alpha }}} } } \right).
 \end{aligned}
\end{equation}
In this paper, we call it Chernoff mutual information relative to Shannon mutual information. \\
In fact, both Shannon mutual information and Chernoff mutual information are  to depict the inner relationship between two random variables. The difference is that they adopt different measures of the distance between two distributions. The relative entropy is adopted in Shannon mutual information and the R{\'{e}}nyi divergence is used in Chernoff mutual information [\cite{van2014renyi}].

\subsection{Outline of the Paper}
The rest of this paper is organized as follows. Section II gives a brief introduction of the several special channel models. In this section, we lists binary symmetric channel (BSC) and multiple symmetric channel (MSC). Furthermore, this section also give a brief introduction of the ACE algorithm. In Section III, some simulated examples are given. Next, Section IV gives the analysis of the simulated examples. Finally, we present the conclusion in Section V.

\section{Channel Information Measures and ACE Algorithm Description}
The mutual information between the input and output of a channel can be used to describe the transmittability of a channel. For example, the maximum mutual information is defined as the channel capacity for a discrete memoryless channel. Obviously, the mutual information is the measure of channel information.\\

\subsection{Binary Symmetric Channel}
\begin{figure}
  \centering
  \includegraphics[width=0.4\textwidth]{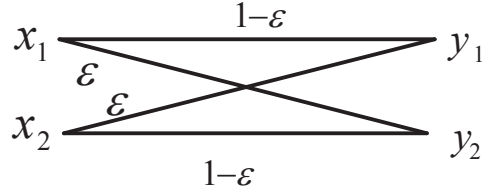} \\
  \caption{Binary symmetric channel.}\label{fig:channelBSC}
\end{figure}

Consider the BSC, which is shown in Fig.\ref{fig:channelBSC}. It is a binary channel where the probability of the input symbols is $(\lambda,1-\lambda)$. The transmission error in it is $\varepsilon $. It is argued in [\cite{Elements}] that it reflects common characteristics of the general channel with errors though it is the simplest model.\\

The Shannon mutual information is given by:

\begin{equation}\label{BSC Mutual information}
\begin{aligned}
 &  I_S(X;Y) = H(Y) - H(Y|X)  \cr
  &  = \sum\limits_{k = 1}^2 {\sum\limits_{j = 1}^2 {p({x_k},{y_j})log{{p({x_k},{y_j})} \over {p({x_k})p({y_j})}}} }   \cr
  &  = (1 - \varepsilon )\log {{1 - \varepsilon } \over {\lambda (1 - \varepsilon ) + (1 - \lambda )\varepsilon }} + \varepsilon \log {\varepsilon  \over {\lambda \varepsilon  + (1 - \lambda )(1 - \varepsilon )}}. \cr
 \end{aligned}
\end{equation}
It can be seen as $I_S(X,Y) = D(p(x,y)||p(x)p(y))$. While the Chernoff mutual information is given by:
\begin{equation}\label{BSC Chernoff information}
\begin{aligned}
 &I_C(X;Y)=- \mathop {\min }\limits_{0 \le \alpha  \le 1} \log \left( {\sum\limits_x {P_1^\alpha (x)P_2^{1 - \alpha }(x)} } \right)\\
 &=- \mathop {\min }\limits_{0 \le \alpha  \le 1} \log (  {(1 - \varepsilon )^\alpha }{\left( {\lambda (1 - \varepsilon ) + (1 - \lambda )\varepsilon } \right)^{1 - \alpha }} \\
 &+ {\varepsilon ^\alpha }{\left( {\lambda \varepsilon  + (1 - \lambda )(1 - \varepsilon )} \right)^{1 - \alpha }} )
  \end{aligned}
\end{equation}
Unfortunately, there is no explicit solution for Eq.(\ref{BSC Chernoff information}). As a result, it is arduous to analyze the Chernoff mutual information.\\
It is observed that the Shannon mutual information seems to be completely different from the Chernoff mutual information.

\subsection{Multiple Symmetric Channel}
\begin{figure}
  \centering
  \includegraphics[width=0.2\textwidth]{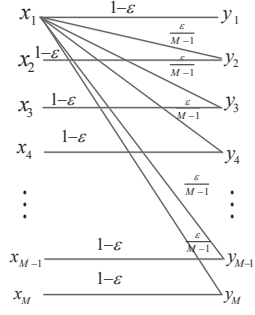} \\
  \caption{Multiple symmetric channel.}\label{fig:channelMBSC}
\end{figure}

The MSC is shown in Fig.\ref{fig:channelMBSC}. The number of input symbol is $M$. The probability of the input symbol is $(\lambda_1,\lambda_2,...,\lambda_M)$ with $\lambda_1+\lambda_2+...+\lambda_M=1$ ($\lambda_i>0$).The transmission error of each one is $\frac{\varepsilon}{M-1}$. The BSC is the special case of MSC for $M=2$.

 Similarly, the Shannon mutual information is given by:
 \begin{equation}\label{MSC Mutual information}
\begin{aligned}
 &  I_S(X;Y) = D(p(x,y)||p(x)p(y)) \cr
  &  = \sum\limits_{k = 1}^M {\sum\limits_{j = 1}^M {p({x_k},{y_j})log{{p({x_k},{y_j})} \over {p({x_k})p({y_j})}}} }   \cr
 \end{aligned}
\end{equation}
The Chernoff mutual information is given by:
\begin{equation}\label{MSC Chernoff information}
\begin{aligned}
 &I_C(X;Y)=- \mathop {\min }\limits_{0 \le \alpha  \le 1} \log \left( {\sum\limits_x {{p(xy)}^\alpha (x){(p(x)p(y))}^{1 - \alpha }(x)} } \right)\\
  \end{aligned}
\end{equation}

\subsection{ACE Algorithm Description}
 Much of research in regression analysis has examined the optimal transformation between one or more predictors and the response. Unlike traditional multiple regression algorithm which requires the priori information of the functional forms, the ACE algorithm of Breiman and Friedman (1985) in [\cite{breiman1985estimating}] does not require that and it is non-parametric transformation. That is to say, it is a fully automated algorithm to estimate the optimal transformation between predictors and response. Furthermore, it can be also used to estimate maximal correlation among random variables. The implementation of the ACE algorithm can consult [\cite{breiman1985estimating}][\cite{wang2004estimating}].

Assume random variables $X_1,X_2,\cdots ,X_p$ are predictors and $Y$ is response. Supposing $\phi _1(X_1),\phi _2(X_2),\cdots,\phi _p(X_p),\theta (Y)$ are arbitrary zero mean functions of the corresponding variables, the residual error is
\begin{equation}
{e^2}\left( {\theta ,{\phi _1},\cdots,{\phi _p}} \right) = {{E\left\{ {{{\left[ {\theta (Y) - \sum\limits_{i = 1}^p {{\phi _i}({X_i})} } \right]}^2}} \right\}} \over {E\left[ {{\theta ^2}(Y)} \right]}}
\end{equation}

\par The algorithm is summarized in Alg.\ref{alg:ACE} as in [\cite{breiman1985estimating}].

\begin{algorithm}[htb]
\caption{ACE Algorithm}
\label{alg:ACE}
\begin{algorithmic}[1]
\STATE set $\theta (Y) = Y/\left\| Y \right\|$ and ${\phi _1}({X_1}),...,{\phi _p}({X_p}) = 0$;\\
\STATE Iterate until $e^2(\theta,\phi_1,...,\phi_p)$fails to decrease;\\
\STATE Iterate until $e^2(\theta,\phi_1,...,\phi_p)$fails to decrease;\\
\STATE For $k=1$ to p Do:\\
\begin{equation*}
 {\phi _{k,1}}({X_k}) = E\left[ {\theta (Y) - \sum\limits_{i \ne k} {{\phi _i}\left( {{X_i}} \right)} \left| {{X_k}} \right.} \right];
\end{equation*}
replace $\phi_k(X_k)$ with $\phi_{k,1}(X_k)$;\\
End For Loop;\\
\STATE End Inner Iteration Loop;\\
\STATE \begin{equation*}
{\theta _1}(Y) = {{E\left[ {\sum\limits_{i = 1} {{\phi _i}({X_i})\left| Y \right.} } \right]} \over {\left\| {E\left[ {\sum\limits_{i = 1}^p {{\phi _i}({X_i})\left| Y \right.} } \right]} \right\|}};
\end{equation*}
replace $\theta(Y)$ with $\theta_1(Y)$;\\
\STATE End Outer Iteration Loop;\\
\STATE $\theta,\phi_1,...,\phi_p$ are the solution $\theta^*,\phi^*_1,...,\phi^*_p$;\\
\end{algorithmic}
\end{algorithm}

In this paper, we use the ACE algorithm help us separate multivariate function. Let $X_1,X_2,\cdots,X_p$ be the channel parameters and $Y$ be measured value of channel information. The functional relation between them is:
\begin{equation}\label{Eq ACE Alg4}
Y = f({X_1},{X_2},...,{X_p}).
\end{equation}
It is supposed that ${X_1},{X_2},...,{X_p}$ are known and independent. Moreover, the functional relation $f$ is also known since it is defined by human to describe the channel information. Therefore, it is easy to get $Y$ and they form a data set $\{Y,X_1,\cdots,X_p\}$. This data set meets the precondition of the ACE algorithm, so ACE algorithm can be used to analyze them. As a result, one can get
\begin{equation}\label{ACE_Def}
 \theta (Y) = \sum\limits_{i = 1}^p {{\phi _i}(X_i)}  + \delta
\end{equation}
where $\delta$ is residual error. In this case, it is easy to find out the separate influence of each correspondent channel parameter.

\section{Simulated Example}
In this section, various numerical simulation results will be presented to analyze the two channel information measures. We focus on conducting the Monte Carlo simulation by computer to compare the Shannon and Chernoff mutual information in different channels by ACE algorithm. The procedure is given by the following simulation procedure.
\begin{algorithm}[htb]
\caption*{Simulation Procedure}
\label{alg:simulated}
\begin{algorithmic}[1]
\STATE Generate channel parameters, such as $\lambda$ and $\varepsilon$;
\STATE Calculate the channel information measures, such as Shannon mutual information;
\STATE Decompose these channel information measures by ACE algorithm;
\STATE Repeat above process till the required times.
\end{algorithmic}
\end{algorithm}

\subsection{Binary Symmetric Channel}
$20000$ observations geneated from the Eq.(\ref{BSC Mutual information}) and Eq.(\ref{BSC Chernoff information}) where $\lambda$ is the probability of the input symbol and $\varepsilon$ is the transmission error. $\lambda$ and $\varepsilon$ are independently drawn from a uniform distribution $U(0,1)$. In this case, our channel information is a multivariate case with two input parameters.

\begin{figure*}{}
  \centering
  \includegraphics[width=0.6\textwidth]{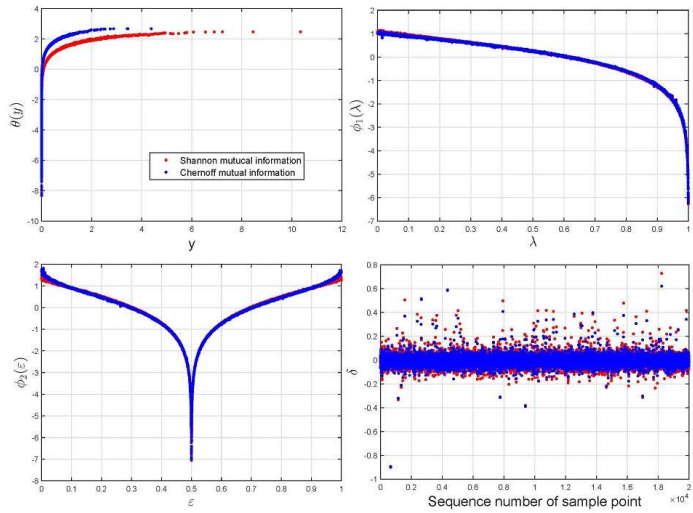} \\
  \caption{ACE optimal transformations of simulated dataset in BSC. The correlation between $\theta(y)$ and $\phi_1(\lambda)+\phi_2(\varepsilon)$ in Shannon mutucal imformation is 0.9994. The same correlation in Chernoff mutucal imformation is 0.9991.}\label{fig:BSC C and I}
\end{figure*}

The ACE algorithm is applied to this simulated data set and the results is shown in Fig.\ref{fig:BSC C and I}. The correlation between $\theta(y)$ and $\phi_1(\lambda)+\phi_2(\varepsilon)$ is extremely close to $1$. Furthermore, The error of the ACE decomposition $\delta$ is invariably near to zero. Clearly, both of them have shown that the ACE decomposition results are excellent. 

It is noted that function curve of $\phi_1(\lambda)$ and $\phi_2(\varepsilon)$ w.r.t. Shannon mutual information is almost coincided with that w.r.t Chernoff mutual information after the ACE decomposition. As a result, in the range of the errors permitted, the $\phi_1(\lambda)$ and $\phi_2(\varepsilon)$ appears to be coincided with each other. $\phi_1(\lambda)$ is monotone decreasing function. The greater $\lambda$, the faster the decline in $\phi_1(\lambda)$. Furthermore, $\phi_2(\varepsilon)$ is symmetric around the line $\varepsilon = 0.5$ for the fact that the channel is symmetrical about $\varepsilon$.

 The results of Shannon mutual information are smaller than the Chernoff mutual information, but whole variant trend is identical. $\theta(y)$ increases rapidly when $y$ is close to zero and the slope of its curve is becoming smaller as $y$ increases.

\subsection{Multiple Symmetric Channel}
$60000$ observations generated from the Eq.(\ref{MSC Mutual information}) and Eq.(\ref{MSC Chernoff information}) where $\lambda_1,\lambda_2,\lambda_3$ is the probability of the input symbol and $\varepsilon$ is the transmission error. They are generated randomly and independently, which is bound in $(0,1)$. In this case, our channel information is a multivariate case with four input parameters.

\begin{figure*}
  \centering
  \includegraphics[width=0.7\textwidth]{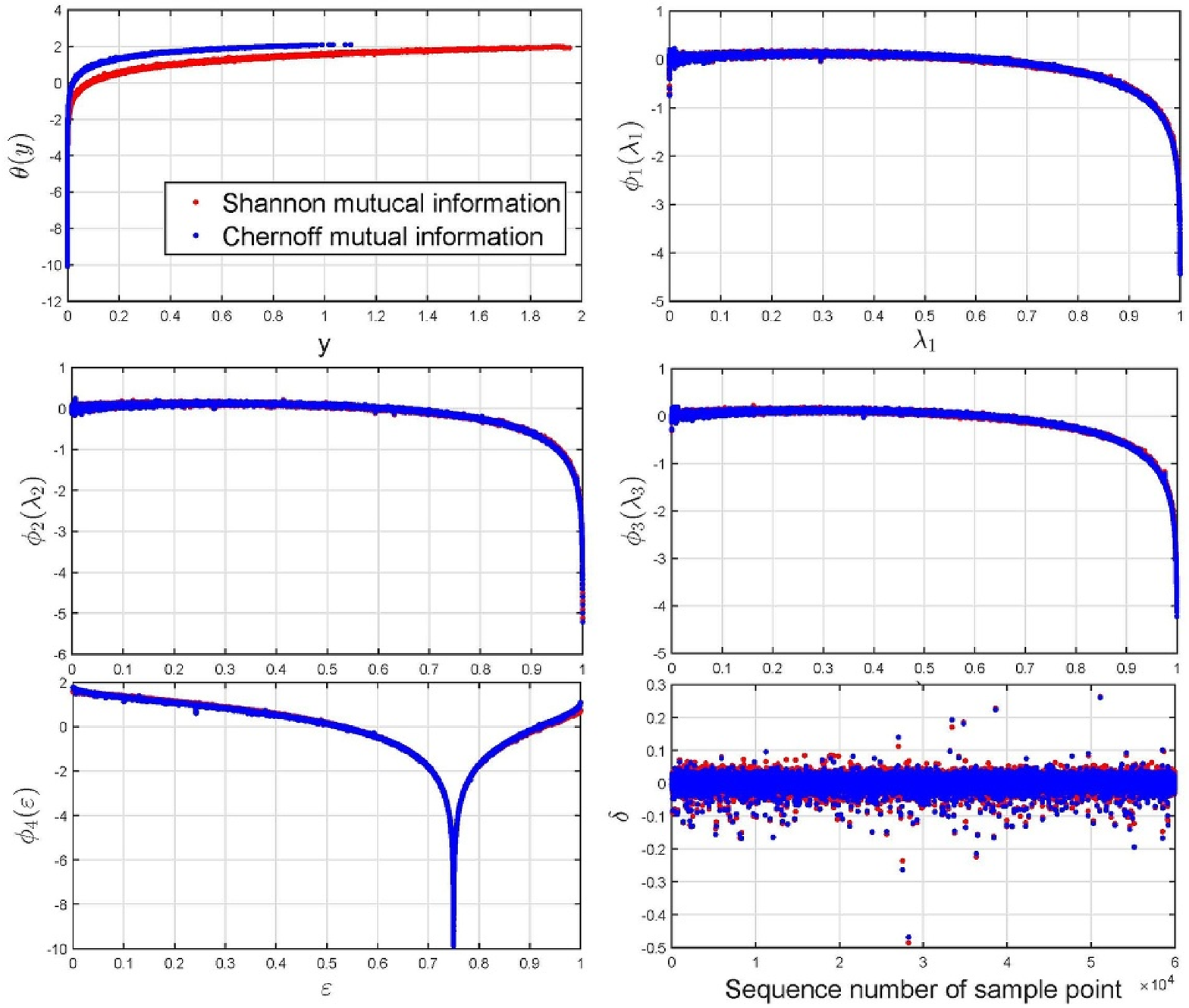} \\
  \caption{ACE optimal transformations of simulated datasetin MSC. The correlation between $\theta(y)$ and $\phi_1(\lambda_1)+\phi_2(\lambda_2)+\phi_3(\lambda_3)+\phi_4(\varepsilon)$ in Shannon mutucal imformation is 0.9999. The same correlation in Chernoff mutucal imformation is 0.9999.}\label{fig:MSC C and I}
\end{figure*}

Fig.\ref{fig:MSC C and I} shows similar characteristics that in Fig.\ref{fig:BSC C and I}. The value of the ACE decomposition error $\delta$ is still very small. On the other hand, the correlation between $\theta(y)$ and $\phi_1(\lambda_1)+\phi_2(\lambda_2)+\phi_3(\lambda_3)+\phi_4(\varepsilon)$ is close to $1$. These two points verify the validity of the ACE algorithm again. 

It is obviously that the function curves of $\theta(y)$,$\phi_1(\lambda_1)$,$\phi_2(\lambda_2)$,$\phi_3(\lambda_3)$ and $\phi_4(\varepsilon)$ are almost the same for these two channel information measures. It is worth noting that the minimum of $\phi_4(\varepsilon)$ occurs when $\varepsilon \approx 0.75$. The values of $\phi_4(\varepsilon)$ is big when $\varepsilon$ is very close to $0$ or $1$. Moreover, $\phi_1(\lambda_1)$,$\phi_2(\lambda_2)$ and $\phi_3(\lambda_3)$ are very similar. Their function curve is flat and close to zero when the independent variable (the probability of the input symbol) is less than $0.7$ . When the independent variable approaches to $1$,  the values of them rapid decrease. In fact, it is reasonable because these input variables are equivalent for the channel.

 As illustrated in Fig.\ref{fig:MSC C and I}, the result of Shannon mutual information is smaller than the Chernoff mutual information, but whole varying trend is identical. $\theta(y)$ increases rapidly when $y$ is also close to zero and the slope of its curve is becoming smaller as $y$ increases.

\section{Discusssion and Physical Explanation}

\subsection{Shannon Shakes Hands with Chernoff}
When more narrowly examined, there are more interesting conclusions. In this section, we analyze these ACE simulated results from the viewpoint of big data. There are several nature behind it. First of all, the data does not lie. Naturally, the ACE results can provide much true and useful information for us. Furthermore, we always turn to numerical analysis when it is difficult to do theoretical analysis. The complicated equations of Shannon or Chernoff mutual information are arduous to do intuitive theory analysis, especially when they are multivariable, so it is quite appropriate to do numerical analysis by ACE algorithm. In fact, the probably approximately correct (PAC) model is enough for us in most of time [\cite{vapnik2006estimation}]. In this model, one only need to construct algorithms which guarantee that it is correct with high probability (not necessarily one). This model is the foundation of Support Vector Machine (SVM) [\cite{vapnik2006estimation}].

\begin{figure}
  \centering
  \includegraphics[width=0.4\textwidth]{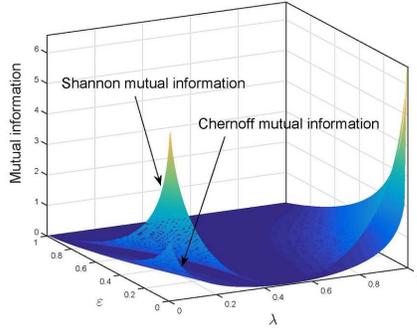} \\
  \caption{The comparison of Shannon mutual information and Chernoff mutual information}\label{fig:BSC1 C and I}
\end{figure}

 The ACE results is
\begin{equation}\label{Res_Fun}
\begin{aligned}
Y = f\left( {\textbf{X}}   \right)  \overset{(a)}{=}{\theta ^{ - 1}}\left( {\sum\limits_{i = 1}^p {{\phi _i}({X_i})} } \right)= \psi \left( {\varphi \left( { {\textbf{X}} } \right)} \right)\\
 \end{aligned}
\end{equation}
where $Y$ is the channel information and ${\textbf{X}}$ is all the set of all the input channel parameters. The equation ($a$) holds because the residual error $\delta$ can be ignored. From the figures above, it is noted that $ \phi_i(X_i)$ is almost the same for both Shannon and Chernoff mutual information. For these two measures, the only difference is $\theta(\cdot)$. The function $\psi(\cdot)$ is different and ${\varphi \left( { {\textbf{X}} } \right)} $ is the same. From the Fig.\ref{fig:BSC C and I} to Fig.\ref{fig:MSC C and I}, it is also noted that the function $\theta(\cdot)$ is monotonic function. As mentioned above, the curves of $\theta(\cdot)$ have the same trend and the values of $\theta(Y)$ for Shannon mutual information is invariably smaller than those for Chernoff information. Therefore, the Shannon's $\theta^{-1}(\cdot)$ is bigger than Chernoff's. Actually, the facts extremely agree with the this corollary, which can well explain the phenomena in Fig.\ref{fig:BSC1 C and I}.

 In this sense, Shannon shakes hands with Chernoff. Over the last decades, Shannon information and Chernoff information are always considered to be two kinds of channel information measures and the relation between them is that Kullback-Leibler divergence is the special case of R{\'{e}}nyi divergence when $\alpha=1$. The channel information measures deduced from them look very different even for the channel as simple as BSC. However, with the aid of the ACE algorithm, we find the inner relations between them. In fact, they are different metrics of the same physical quantity. For a channel, in a certain time, if all of its parameters are given, all of its state and properties will be determined. Thus the information about the same channel is always consistent. For example, the kilometer and centimeter are both measures of length, but they are used in different situations for convenience. Generally speaking, we do not adopt the centimeter to measure the distance between two cities. In contrast, we would like adopt it to measure the length of the steel rule. Their true values are identical when measuring the same things, even their numerical values are different. Similarly, the Shannon and Chernoff information are two metrics of the same quantity of information.

\subsection{The Nature of Channel Information}
Obviously, the ability to convey information of a channel is decided by input parameters. That is to say, the parameters represent the channel. As a result, we hold opinion that these channel parameters decide the nature of channel. It is noted that $\varphi \left(  {\textbf{X}}\right)$ in Eq.(\ref{Res_Fun}) is the same whatever we adopt the Shannon channel information measures or Chernoff channel information measures. In this paper, we take $\varphi \left(  {\textbf{X}}\right)$ as the nature of channel information and we get
\begin{equation}\label{Nature_Fun}
\begin{aligned}
\varPsi=\varphi \left( {\textbf{X}} \right)
 \end{aligned}
\end{equation}
where the $\varPsi$ is the nature of channel information. It is only decided by the channel parameters and it contains the core of the channel information. The other physical quantity about channel information is just the function of $\varPsi$, just as
\begin{equation}\label{Nature_Fun1}
\begin{aligned}
  \rm{I}= \psi (\varPsi)
 \end{aligned}
\end{equation}
where $\rm{I}$ is a measurement of channel information. As far as Shannon mutual information and Chernoff mutual information are concerned, their only difference is the function $\psi(\cdot)$. They can be seen as a compound function. The inner is $\varPsi$ which is the function w.r.t. channel parameters and the outer is the function w.r.t. $\varPsi$. For different channel information measures, the inner is always the same but the outer is different.

There is a conjecture that if a novel information measure is put forward to measure the channel information, it will be decomposed to a function of $\psi(\cdot)$ by ACE algorithm from a very large extent. This rule can be used to judge the rationality of the new information measure. This conjecture also provides a way for us to propose a new information measure for the fact that we can construct that based on designing the function $\psi(\cdot)$. In fact, even if this guess is not right with probability one, the new one which conforms to the law is correct with large probability and we can easily find the relation between that and previous information measures, such as Shannon and Chernoff information.

\section{Conclusion}
In this paper, the different channel information measures in several channels are revisited in Big Data viewpoint by ACE algorithm. Fortunately, the decomposition results of independent variable is the same with a slightly difference in the function of channel information values. That is to say, Shannon shakes hands with Chernoff with the fact that they are just two metrics of the same quantity of channel information. For every channel, there is a nature of channel information, which is only decided by the channel parameters no matter what information measures are adopted. In fact, the Big Data viewpoint such as ACE algorithm provides a new viewpoint for us to reexamine the Information Theory and find out that Shannon and Chernoff shake hands with each other, which have been ignored for decades. Furthermore, this method can help us to construct new information measures with keeping the nature of the channel information unchanged. Obviously, we put forward a criterion to judge whether a new channel information measure be appropriate.

\section*{Acknowledgement}
This work was supported by the China Major State Basic Research Development Program (973 Program) No.2012CB316100(2) and National Natural Science Foundation of China (NSFC) NO.61171064.

{\small
\bibliographystyle{IEEEtran}

}

\end{document}